\documentclass[a4paper]{PoS}
\usepackage{amsmath}
\pdfoutput=1

\title{ Measurement of Transverse Single Spin Asymmetries in $\pi^0$ Production from
$p^{\uparrow}+p$ and $p^{\uparrow}+A$ Collisions at STAR }

\ShortTitle{$A_N$ of forward $\pi^0$s in $p+Au$ collisions at STAR}

\author{
  \speaker{Christopher Dilks}
  \\
  Pennsylvania State University\\
  E-mail: \email{cjd5150@psu.edu}
}

\abstract{
In 2015 the first collisions between polarized protons and nuclei occurred at the
Relativistic Heavy Ion Collider (RHIC), at a center-of-mass energy of $\sqrt{s_{NN}}=200$ GeV.
Comparisons between spin asymmetries and cross-sections in $p+p$ production to those in $p+A$
production provide insight into nuclear structure, namely nuclear modification factors, nuclear
dependence of spin asymmetries, and comparison to models with saturation effects.  The transverse
single-spin asymmetry, $A_{N}$, has been measured in $\pi^{0}$ production in the STAR Forward Meson
Spectrometer (FMS), an electromagnetic calorimeter covering a forward psuedorapidity range of
$2.6<\eta<4$. Within this kinematic range, STAR has previously reported the persistence of large
$\pi^0$ asymmetries with unexpected dependences on $p_T$ and event topology in $p+p$ collisions.
This talk will compare these dependences to those in $p+A$ production.
}

\FullConference{XXIV International Workshop on Deep-Inelastic Scattering and Related Subjects\\
		11-15 April, 2016\\
		DESY Hamburg, Germany}

\begin{document}

\section{Introduction} 

One of the principal observables which helps to gain insight on the spin and
orbital angular momenta of partons within polarized protons is the transverse single-spin asymmetry,
denoted $A_N$. Letting $d\sigma^{\uparrow\left(\downarrow\right)}$ denote the differential
cross-section for leftward scattering of a spin up (down) polarized proton on an unpolarized target,
$A_N$ is defined as the following ratio: 
\begin{equation}
\label{tssa}
A_N:=\displaystyle\frac{d\sigma^{\uparrow}-d\sigma^{\downarrow}} {d\sigma^{\uparrow}+d\sigma^{\downarrow}}.
\end{equation}
This asymmetry shows up as a $\cos\phi$ modulation of the cross-section, where $\phi$ is the azimuth
of the observed particle. 
Since 1976, large values of $A_N$ have been seen in forward pion production, with only modest
dependence on the center-of-mass energy, $\sqrt{s}$ \cite{klem}. For forward $\pi^0$s, $A_N$ increases with
respect to $x_F:=2p_z/\sqrt{s}$, where $p_z$ is the longitudinal momentum of the $\pi^0$, and $A_N$
is mostly flat or rising with respect to $\pi^0$ transverse momentum, $p_T$.  
\cite{star2008}\cite{heppelmannAN}\cite{mondalAN}.

In 2015, the world's first polarized $p^{\uparrow}+A$ synchrotron collisions took place at the
Relativistic Heavy Ion Collider (RHIC), at $\sqrt{s}=200$ GeV.
Significant data samples of polarized protons colliding against gold ($A=197$) and aluminum ($A=27$)
nuclei were obtained. With these new data, there are a few observables and possible implications
available. The first natural question from a spin physics point of view is how $A_N$ in
$p^{\uparrow}+A$ compares to that in $p^{\uparrow}+p$ collisions. One implication of this
comparison is its use as a possible probe of gluon saturation: the color glass condensate model
predicts that $A_N$ in $p^{\uparrow}+A$ decreases as $A$ increases \cite{cgcmodel}. The nuclear
modification factor, $R_{pA}$, and its kinematic-dependences can also be assessed. Dependences of
$A_N$ on event topology can also be compared between $p^{\uparrow}+A$ and $p^{\uparrow}+p$
collisions, possibly leading to insight on fragmentation universality. Finally, dependences of $A_N$
as well as $R_{pA}$ on collision centrality is another avenue for exploring possible effects a
nuclear medium has on these observables.

\section{Forward Calorimetry at STAR}

\begin{figure}[h!]
\centerline{\includegraphics[width=0.5\textwidth]{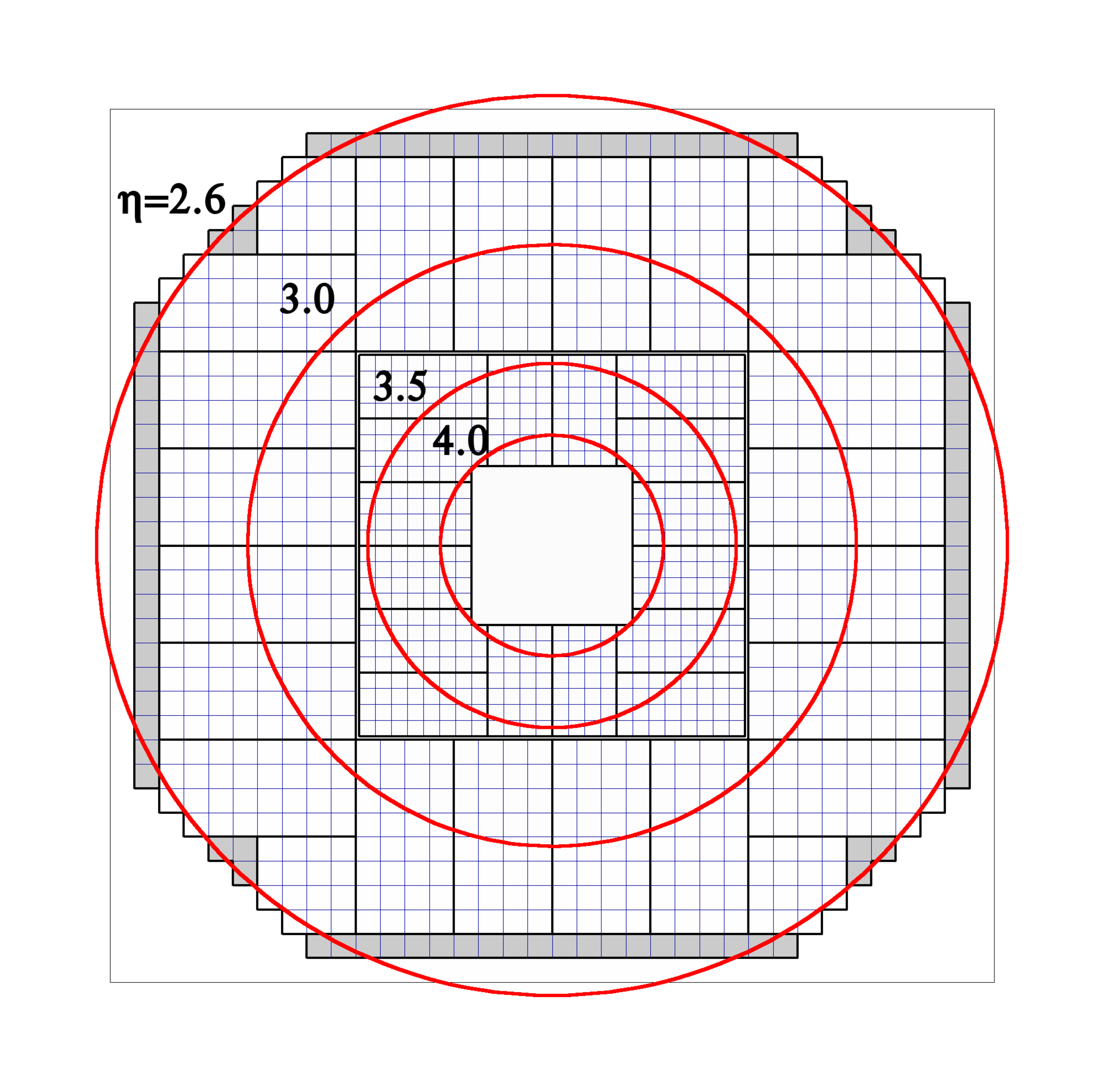}}
\caption{Schematic of the FMS Pb-glass cells. Red circles indicate
rings of constant $\eta$, bold black
double-lines indicate the boundary between large outer cells and small inner cells, and grey-colored
cells are not in the trigger. The beam-pipe passes through the white square in the center.} 
\label{FMSschematic}
\end{figure}

The primary detector used in this analysis is the Forward Meson Spectrometer (FMS), a Pb-glass
electromagnetic calorimeter covering a pseudorapidity range of $2.6<\eta<4$. It is composed of 1,264
Pb-glass cells, each coupled to photomultiplier tubes, arranged in a square array as shown in
Figure \ref{FMSschematic}. The primary observable is the $\pi^0\to\gamma\gamma$ decay channel, where
the transverse distance between the two photons along with their energies provide access to the
di-photon invariant mass, $M_{\gamma\gamma}$, and subsequently to $\pi^0$ reconstruction. 

\begin{figure}[h!]
\centerline{\includegraphics[width=0.75\textwidth]{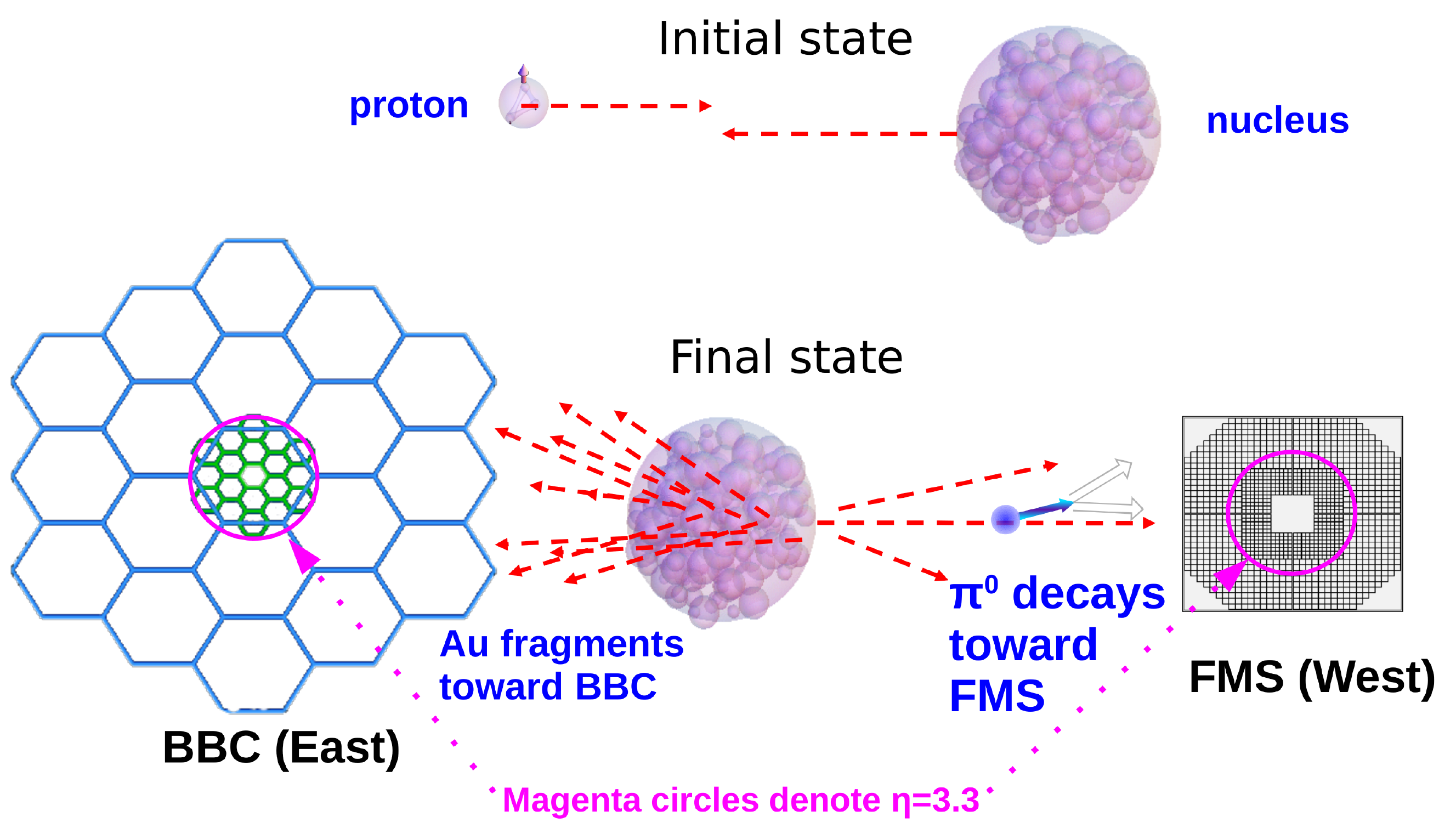}}
\caption{Schematic of the $p+A$ experimental setup, with the FMS and BBC shown in the final-state
setup. The proton beam faces the FMS and the nuclear beam faces the opposite $\eta<0$ BBC. Only the
scintillators inside the pink $\eta=3.3$ ring are considered in this analysis; this $\eta$-ring is
also drawn on the FMS schematic for comparison.} 
\label{BBCschematic}
\end{figure}

Another detector, which is used in quantifying systematic uncertainties in this analysis, is the
pair of Beam Beam Counters (BBCs). Each BBC is an annular arrangement of hexagonally-tiled
scintillator panels, as shown in the left side of Figure \ref{BBCschematic}. The portion of the BBC
used in this analysis covers a pseudorapidity range of $3.3<\left|\eta\right|<5$, where $\eta>0$
denotes forward production ({\it i.e.}, toward the FMS) and $\eta<0$ denotes backward production
({\it i.e.}, opposite the FMS). In $p+A$ collisions, the proton beam faces the FMS and the nuclear
beam remnants scatter toward $\eta<0$. 

In order to reconstruct $\pi^0$s, photons within 35 mrad isolation cones were collected. Within each
event, only the highest energy photon pair in the highest energy isolation cone cluster was
considered.  This event selection yielded a $M_{\gamma\gamma}$ distribution with a lower
background than less-restrictive event selections. To select $\pi^0$s from di-photon events, a mass cut of
$\left|M_{\gamma\gamma}-135\right|<120~\text{~MeV}/c^2$ was used, along with an energy-sharing cut
of $\left|E_1-E_2\right|/\left(E_1+E_2\right)<0.7$, where $E_{1,2}$ denote the energy of the
photons. Finally, after demanding the $\pi^0$ $p_T$ to be above trigger threshold, these $\pi^0$s
were then organized into $E$ and $p_T$ bins for subsequent asymmetry analysis. 

The selected $\pi^0$s, within their $E$ and $p_T$ bins, are then binned in $\cos\phi$ bins.
Letting $N^{\uparrow\left(\downarrow\right)}$ denote the $\pi^0$ yield from the scattering of
spin-up(down) protons, 
the distribution of the raw transverse single-spin asymmetry is 
fit to the following linear equation with parameters $p_{0,1}$:
\begin{equation}
\label{linear_fit}
\dfrac{N^{\uparrow}-N^{\downarrow}}{N^{\uparrow}+N^{\downarrow}}=
p_0+p_1\cos\phi,
\end{equation}
The measured $A_N$ is then the amplitude of the azimuthal modulation of this
quantity, divided by the beam polarization $P$, that is, $A_N=p_1/P$, whereas $p_0$ scales with the
relative luminosity between spin up and spin down protons.

In $p+A$ collisions (and to a lesser extend in $p+p$ collisions), the extracted $A_N$ depends on the
away-side BBC multiplicity, {\it i.e.}, on the charged particle distribution from the nuclear
remnants. In this analysis, this dependence is characterized as the dominant systematic uncertainty
on $A_N$ within each kinematic bin; further characterization of this dependence and its possible
relation to centrality will be studied in future analyses.

\section{Results and Discussion} 
\begin{figure}[h!]
\centerline{\includegraphics[width=\textwidth]{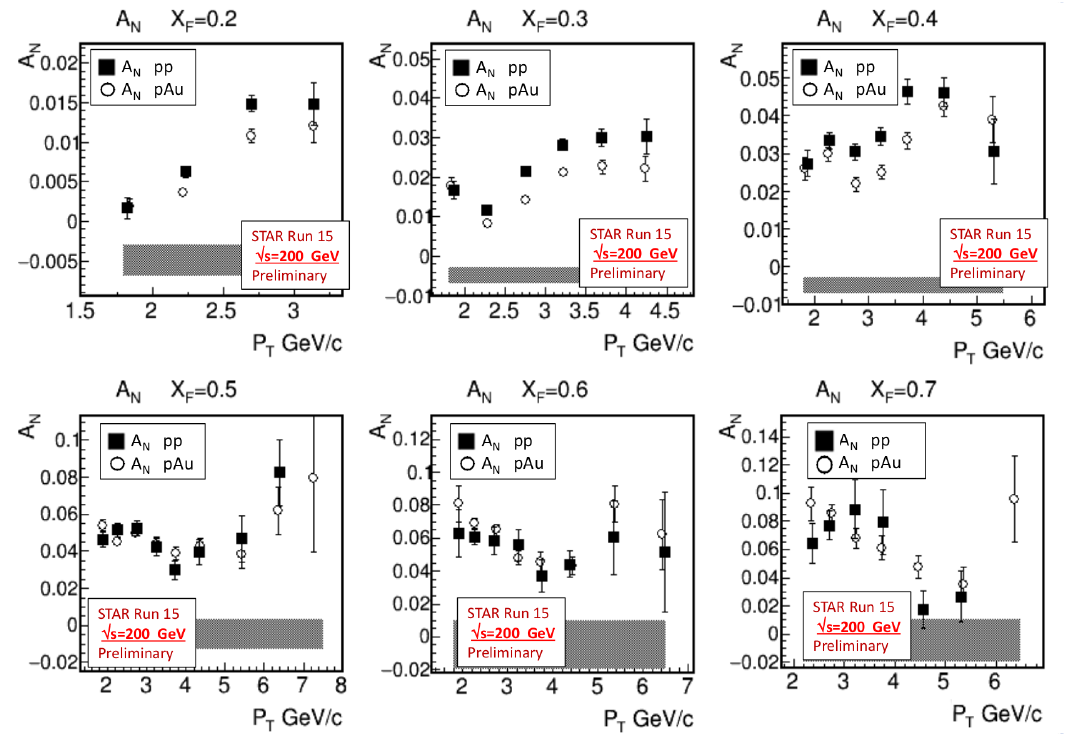}}
\caption{Transverse single-spin asymmetry $A_N$ vs. $p_T$ for $\pi^0$s from $p+p$ (filled points)
compared to $p+Au$ (open points) for six $x_F$ bins of widths 0.1 centered about the value indicated
in the plot titles. Vertical error bars are statistical uncertainties and the vertical size of the
grey band centered around $A_N=-0.005$ is the size of the systematic uncertainty from the BBC
multiplicity. (Note: vertical scales are different in each plot)} 
\label{A_N}
\end{figure}

The transverse single-spin asymmetries for forward $\pi^0$s
produced in $p+p$ and in $p+Au$ collisions are shown as a function of $p_T$ in six different
$x_F\approx E_{\pi^0}/(100~\text{~GeV})$ bins in Figure \ref{A_N}, where the $x_F$ approximation is
valid for forward production kinematics; filled-in points are for $\pi^0$s from $p+p$ and
open points are those from $p+Au$. Statistical uncertainties are represented by vertical error bars
and the dominant systematic uncertainty from the charged particle multiplicity in the gold-going BBC
is represented by the vertical size of the grey band centered at $A_N=-0.005$. The asymmetry from
$p+p$ is similar to that from $p+Au$, within statistical and systematic uncertainties. These data
represent a luminosity of approximately $35~\text{pb}^{-1}$ from $p+p$ and $205~\text{nb}^{-1}$
from $p+Au$. Forward proton beam average polarizations were $55.6\pm 2\%$ and $60.4\pm 2\%$ for $p+p$ and
for $p+Au$, respectively.

\begin{figure}[ht!]
\centerline{\includegraphics[width=0.85\textwidth]{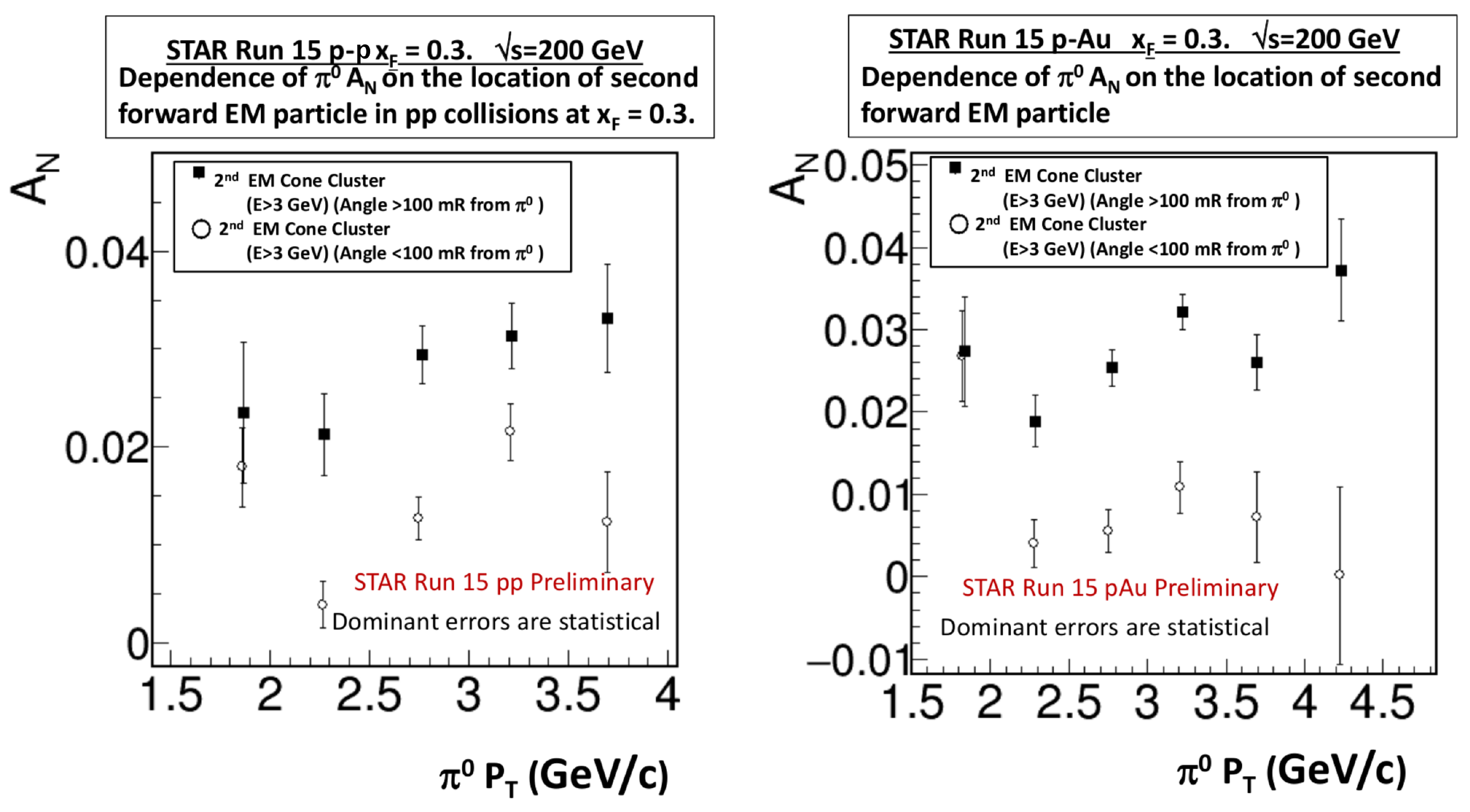}}
\caption{$A_N$ for isolated (filled squares) and non-isolated (open circles) $\pi^0$s from $p+p$
(left) and $p+Au$ (right). (Note: vertical scales are different in each plot)} 
\label{topology}
\end{figure}

Another study that was performed with data from the 2015 RHIC run is the dependence of $\pi^0$
$A_N$ on event topology. Recent data from the FMS in previous RHIC runs have shown that isolated
$\pi^0$s have a higher $A_N$ than those which are surrounded by other forms of electromagnetic
energy \cite{heppelmannAN}\cite{mondalAN}. This dependence was studied again in the 2015 $p+p$ data and was
compared to the $p+Au$ sample. For a given $\pi^0$ in a 35 mrad isolation cone, the next highest
energy photon cluster with $E>3$ GeV outside the isolation cone was considered; its
$\left(\eta,\phi\right)$-position relative to that of the $\pi^0$ was denoted
$\left(\Delta\eta,\Delta\phi\right)$. 
The $\pi^0$ events were then separated into
two isolation classes: those with electromagnetic energy deposited $\Delta\phi<100$ mrad from the
$\pi^0$ and its more-isolated complement, $\Delta\phi>100$ mrad. Figure \ref{topology} shows the
$A_N$ for the two event classes, where more-isolated $\pi^0s$ are plotted as filled-in squares and
less-isolated as open circles. The same trend persists in $p+Au$: more isolated $\pi^0s$ have a
higher $A_N$.

\newpage
 
\newpage
\section{Conclusion}

This analysis was of $A_N$ from $\pi^0$s from the world's first synchrotron collisions of polarized
protons against nuclei at $\sqrt{s}=200$ GeV. The value of $A_N$ in $p+p\to\pi^0+X$ was shown to be
within statistical and systematic uncertainties of that in $p+Au\to\pi^0+X$ for all kinematic bins
analyzed. The value of $A_N$ in both cases appears to be flat or to be rising with respect to $p_T$ and
also rises with $x_F$. Its dependence on event topology is also similar for both cases: isolated
$\pi^0$s are associated with higher values of $A_N$. Plans for future analyses include
characterizing more carefully the dependence of $A_N$ on nuclear breakup multiplicity and its
possible relation to collision centrality, as well as studies of nuclear modification factors
within these kinematics.

\end{document}